\documentstyle[aps,12pt,preprint]{revtex}
\tightenlines
\begin{document}

\newcommand{\nc}{\newcommand}
\newcommand{\BE}{\begin{equation}}
\newcommand{\EE}{\end{equation}}
\newcommand{\BA}{\begin{eqnarray}}
\newcommand{\EA}{\end{eqnarray}}
\newcommand{\V}[1]{\mbox{\boldmath $#1$}}

\title{Electromagnetic Spectrum from QGP Fluid \footnote{
A talk given at the International School on the Physics of
Quark Gluon Plasma, June 3-6, 1997, Hiroshima, Japan.
To be appeared in Prog. Theor. Phys. Supplement.}}
\author{Tetsufumi Hirano$^1$  \thanks{Electronic address :
697l5212@mn.waseda.ac.jp}, Shin Muroya$^2$
\thanks{Electronic address : muroya@yukawa.kyoto-u.ac.jp},
and Mikio Namiki$^1$ \thanks{Electronic address :
namiki@mn.waseda.ac.jp}}

\address{$^1$Department of Physics, Waseda University\\Tokyo
169, Japan}
\address{$^2$ Tokuyama Women's College\\Tokuyama, Yamaguchi
754, Japan}
\date{\today}

\preprint{WU-HEP-97-5, TWC-97-2}
\maketitle
\begin{abstract}

We calculate thermal photon and
electron pair
distribution from hot QCD matter produced in high 
energy heavy-ion collisions, based on a hydrodynamical
model which is so tuned as to reproduce the 
recent experimental data at CERN SPS, and compare these
electromagnetic spectra with experimental data given by CERN WA80
and CERES.
We investigate mainly the effects of the off-shell
properties of the source 
particles on the electromagnetic spectra.

\end{abstract}
\newpage
\section{Introduction}

In high-energy heavy-ion collisions, so many kinds of 
secondary particles, such as hadrons, leptons and photons,
come out of hot matter which is produced at an early stage of
the nuclear reaction.
At first, we naively expect that the particle distribution
directly reflects informations concerning the hot matter. As for
hadrons, however, we regret to say that the final 
distribution is far from the initial form of hot matter 
and it is dirty due to strong interaction. While photons and
leptons are considered to keep the information about the hot
matter, because they interact only electro-magnetically.

We analyze the thermal photon and electron pair
emission from the hot matter in a consistent manner, as
follows:
first we choose initial parameters in our hydrodynamical
model so as to fit the experimental data of hadrons.
Next, we derive the thermal production rates of photon and
dilepton based on a quantum Langevin equation.
Finally, accumulating the production rate over the whole
space-time volume, which is given by our hydrodynamical
model, we obtain the electromagnetic spectra which are to
be compared with experimental data.

\section{Relativistic hydrodynamical model and hadron spectrum}

We first analyze recent hadron spectra at CERN SPS energy.
In a previous paper \cite{Akase91}, we solved numerically 
the (3+1)-dimensional hydrodynamical equation.
In this paper we adopt the Bag model for the
equation of state instead of the simple phase transition
model discussed in Ref.~\cite{Akase91}.
Here we suppose that the fluid in the QGP phase is
composed dominantly of $u$-, $d$-, $s$-quarks and gluons and 
that the fluid in the hadron phase is composed dominantly of 
pions and kaons. We fix the critical temperature as $T_c = 160$ 
MeV and the freeze-out temperature as $T_f = 140$ MeV.

Putting the initial temperature as $T_i = 209$ MeV, 
we can see that our hydrodynamical model well reproduce the
experimental
data of the pseudo-rapidity distribution of charged hadron
and the transverse momentum distribution of neutral pions in
the S+Au 200A GeV collisions \cite{Muroya98}.
In the case of the Pb+Pb 158A
GeV collisions, the best value to 
reproduce the recent experimental data of the rapidity
distribution of negative charged hadron and the transverse
momentum distribution of neutral pions is 
$T_i = 190$ MeV \cite{Muroya98}.

\section{Thermal production rate}

As for the thermal photon and electron pair emission
processes from hot matter,
we can easily write
general formulas of production rates for these
processes \cite{McLerran}
\begin{eqnarray}
R_{\gamma}  & = & \int \frac{d^3 {\V k}}{(2 \pi)^3 2 k}
\sum_{\lambda} \varepsilon_\mu
^{(\lambda)} \varepsilon_\nu ^{(\lambda)}
H_{\mu \nu}(k), \\
R_{e^- e^+} & = & \int \frac{d^3 {\V p}_{e^-}}{(2 \pi)^3 2
\varepsilon_{e^-}} \frac{d^3 {\V p}_{e^+}}{(2 \pi)^3 2
\varepsilon_{e^+}} \nonumber \\
 & \times & e^2 \mbox{Tr}[\gamma^\mu (/
\hspace{-0.2cm} p_{e^-}-m_e) \gamma^\nu(/
\hspace{-0.2cm} p_{e^+}+m_e)]
\frac{1}{P^4}  H_{\mu \nu}(P).
\end{eqnarray}
Here $H_{\mu \nu}$ is a hadronic structure tensor, which is
the Fourier transform of current correlation function.
Supposing that a local equilibrium system is dominated by 
a certain mode and that the canonical 
operator of the
mode obeys a quantum Langevin equation \cite{Mizutani88},
we replace the correlation function with the
ensemble average in the sense of the quantum Langevin equation.
Then we easily obtain the thermal photon production 
rate \cite{Hirano97}.

We can also obtain the thermal dilepton
production rate based on the quantum Langevin equation in the
same manner.
We first consider the process of $\pi^+ \pi^- \rightarrow
e^- e^+$.
The production rate of this process at $T = 160$ MeV
is shown in Fig.~1.
It is well known that this elementary process has the
threshold at $M = 2 m_\pi$,
which is originated from the
on-shell condition of the source particle. 
In our formalism the
source particle 
has off-shell properties, hence, 
we can observe in Fig.~1 that there is no threshold effect.  
Assuming the Vector Meson Dominance, we multiply the 
production rate by the square of pion electromagnetic
form factor.
In Fig.~1,  
the deviation between the usual on-shell model and the
off-shell model appears around the threshold and 
the difference is very small in the invariant mass region
which is much larger than the threshold. 
On the other hand, because of the small value of threshold,
we can safely neglect the off-shell effect in the process
of  $q \bar{q} \rightarrow e^- e^+$.

In the mixed phase region,
We define the production rate as a function of $T$
and $\lambda$ 
\begin{equation}
\frac{dN}{dx^4} = R(T,\lambda) = \lambda R _{\rm{QGP}}
+ (1-\lambda) R_{\rm{had}},
\end{equation}
where $\lambda$ is the fraction of QGP phase region.
Integrating this production rate over the whole space-time 
volume in which the particle source exists, we obtain the
momentum distribution of photon and the invariant mass
distribution of electron pair which are to be compared
with experimental data.

\section{Results}

Using together with the numerical results of our
hydrodynamical model and the thermal production rate, we can 
predict the electromagnetic spectra.
Figure 2 shows the numerical result of thermal
photon distribution in S+Au 200A GeV collisions 
in comparison with the CERN WA80 data \cite{96WA80}.
We can see in this figure
that our model is consistent with these experimental data.
We can predict the thermal photon distribution for the Pb+Pb
158A GeV collisions (Fig.~3).
In Fig.~3, the photon distribution for the 
Pb+Pb collisions 
is similar to the result of the S+Au collisions. The 
 larger space-time volume in 
the Pb+Pb collisions is compensated by the lower initial
temperature in comparison with the S+Au experiment.

Figure 4 shows the numerical results of electron pair
distribution in comparison with the experimental 
data given by CERN CERES \cite{CERES}. The 
solid curve
stands for the electron pair distribution of 
our model with $ c = 1.0 $ in the damping function and
the dashed curve stands for the result
obtained by making use of the production rate based
on the perturbative method 
neglecting off-shell property of the source current. 
The parameter $c$ in the damping function has already been
fixed by the previous
analyses of the thermal photon production \cite{Hirano97}.

Our model based on the quantum
Langevin equation was expected to enable us to reproduce 
the enhancement of the experimental data near 
the threshold $M \sim 2
m_\pi$ because of the off-shell properties of the
source current (see also Fig.~1).
However, we must say from Fig.~4 that 
the off-shell property of the source particles is not enough
to reproduce the enhancement in the small invariant mass
region of the experimental data \cite{Sollfrank}.
In this paper, we do not take account of the possibility of
the partial restoration of chiral symmetry. Therefore,
our results may be improved by taking into account the 
mass shift of $\rho$ meson due to finite temperature effects.

\section{Summary}

We compared our numerical results of electromagnetic spectrum
with experimental data at CERN SPS energy. In the photon case
our result is almost consistent with WA80 S+Au data.
Furthermore, we evaluated the expected thermal photon 
distribution
for the Pb+Pb collisions. We observed that this distribution
is similar to the result of the S+Au collisions.
In the case of the low mass region of electron pair distribution,
our result was not improved by the off-shell property enough
to explain the experimental data obtained by CERN CERES.

\begin{center}
{\bf Acknowledgments}
\end{center}

The authors are much indebted to Prof.\ I.~Ohba 
and Prof.\ H.~Nakazato for
their helpful comments.
They also thank Dr.\ H.~Nakamura, Dr.\ C.~Nonaka 
and other members of 
high energy physics group of Waseda Univ.\ for their
fruitful discussions.


\begin{Large}

FIGURE CAPTION

\end{Large}

FIG.\,1 The production rate of the process
$\pi^+ \pi^- \rightarrow e^- e^+$.
The curve $c = 0$ is the same result as the 
perturbation theory.

FIG.\,2 Thermal photon distribution in the S+Au 200A GeV
collisions. The solid curve stands for the photon
distribution of our numerical result.

FIG.\,3 Expected thermal photon distribution. The solid
curve stands for the expected thermal photon spectrum in 
the Pb+Pb 158A GeV collisions. We also represent the
result of S+Au collisions for comparison.

FIG.\,4 Electron pair invariant mass distribution.

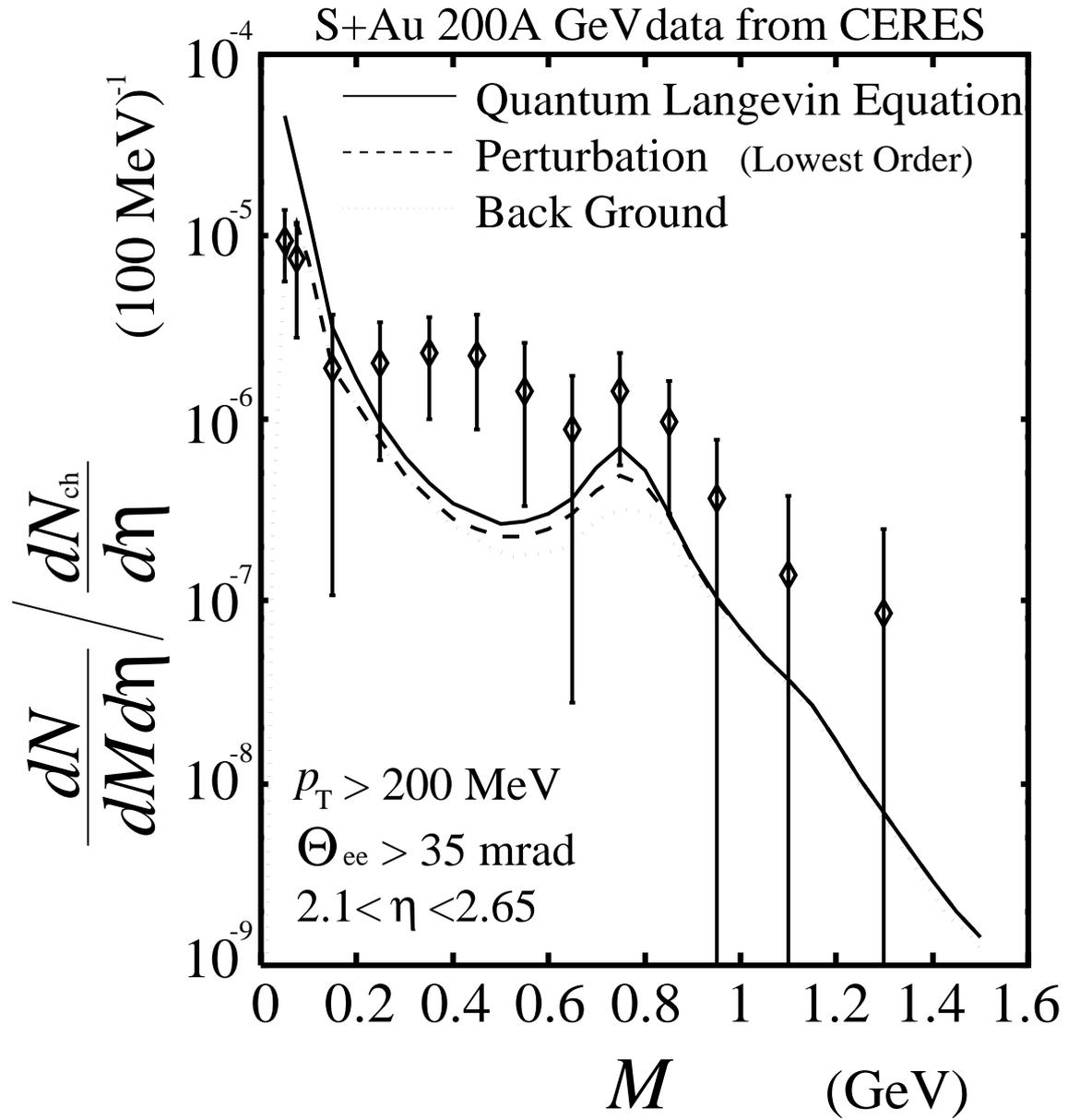
\end{document}